\documentstyle[aps,prl,twocolumn]{revtex}
\begin{document}
\draft

\narrowtext

\noindent
{\large \bf Stability of the Period-Doubled Core of the 90$^\circ$
Partial in Silicon}

\vskip 1mm

In a recent Letter [1], Lehto and \"Oberg (LO) investigated the
effects of strain fields on the core structure of the 90$^\circ$
partial dislocation in silicon, especially the influence of the choice
of supercell periodic boundary conditions in theoretical simulations.
Specifically, they addressed the relative stability of the
traditionally accepted single-period (SP) geometry vs.~our recently
proposed double-period (DP) structure [2].  Performing supercell
calculations that employed the original Keating potential for Si [3],
they reached two main conclusions: (1) that a dipole-cell
configuration minimizes the overall cell strain, thus being more
adequate for small-cell simulations; (2) that the relative stability
of the SP and DP geometries depends on the choice of boundary
conditions, with the SP and DP cores being favored for ``dipole'' and
``quadrupole'' configurations, respectively. The purpose of the
present comment is not to dispute their first point, with which we
agree. Rather, we wish to focus on the more important DP-versus-SP
stability issue, because we believe their second conclusion to be
incorrect. Below, we show that their results for the relative
stability between the two structures are in disagreement with
cell-size converged tight-binding total energy (TBTE) calculations,
which suggest {\it the DP core to be more stable, regardless of the
choice of boundary condition}. Moreover, we argue that this
disagreement is due to their use of a Keating potential.

Clearly, in the limit of sufficiently large supercells, all results
should be independent of the the choice of boundary conditions. Here,
supercell convergence is investigated by performing TBTE and
Keating-potential calculations for three different supercell sizes,
using both the dipole and quadrupole cells, and two different sets of
parameters for the Keating-potential calculations [3,4]. In our larger
cell, all dislocation separations are similar to those of the
2048-atom cells studied by LO. The first observation  we can draw from
the TBTE results in Table~I is that, as pointed out by LO, for small
cells the dipole boundary condition gives results which are closer to
the converged value, while the quadrupole cell has a bias of $\sim$10
meV/\AA\ in favor of the DP structure, which decreases by one order of
magnitude as we approach cell-size convergence. However, Table~I also
shows that the Keating potential has a much stronger bias in favor of the SP
structure. For the 192-atom cell, the results of LDA (from Ref.~[2]),
TETB, and the Keating potentials of Refs.~[3] and [4] for
$E_{DP}-E_{SP}$ are $-$80, $-$70, $-$20, and $-$15, respectively (in meV/\AA,
with error bars of $\sim\!\pm 5$ in each case). Thus,
the Keating potential has a
systematic bias in favor of SP of $\sim$50 meV/\AA\ relative to the
more accurate methods.  As it happens, going to cell-size convergence
shifts all the $E_{DP}-E_{SP}$ values by roughly 20 meV/\AA. As a
result, the converged values for the Keating potentials are close to
zero, making the variations with choice of boundary
conditions (which, in absolute numbers, are similar to
those seen in TBTE) take on an artificial importance.
In fact, the $E_{DP}-E_{SP}$ values are so small that even the
qualitative conclusions about which structure is favored can be seen
to depend on the choice of Keating parameters.
On the other hand, it seems clear that the more accurate TBTE results
would be quite immune from displaying a sign change in $E_{DP}-E_{SP}$ as a
result of details of the choice of cell size or boundary conditions.


Finally, LO also based their conclusions in part on a
density-functional calculation for a finite cylindrical sample.
However, we suggest that the systematic errors involved in choosing
surface boundary conditions for such a geometry may also be severe,
and that such results should not be trusted in the absence of careful
tests of convergence with respect to sample size.
\vspace{-0.1in}
\begin{table}
\caption{Energy of the DP relative to the SP core, in meV/\AA\ per
dislocation, for the 90$^\circ$ partial in Si. Four different
approximations (LDA, TBTE and two different Keating parameterizations)
are used to compare quadrupole and dipole boundary conditions.
Cell size refers to the DP case.}
\begin{tabular}{lrrr}
Cell size (atoms) &192 &576 &1920\\
\hline
LDA\\
\quad quadrupole &$-$79  & & \\
TETB\\
\quad quadrupole &$-$74  &$-$63 &$-$55 \\
\quad dipole     &$-$62  &$-$55 &$-$55 \\
Keating, Ref.~[3]\\
\quad quadrupole &$-$27  & $-$7  &  1 \\
\quad dipole     &$-$14  &  2    &  5 \\
Keating, Ref.~[4]\\
\quad quadrupole &$-$22  &$-$10  & $-$5 \\
\quad dipole     &$-$13  & $-$5  & $-$3 \\
\end{tabular}
\end{table}
R. W. Nunes acknowledges financial support from CNPq-Brazil.
\vskip 1mm
\noindent R. W. Nunes$^1$ and David Vanderbilt$^2$\par
$\;^1$ Depto.\ de F\'{\i}sica, UFMG
\par$\;\;\;$  CP~702, Belo Horizonte, MG 30123-970, Brazil\par
$\;^2$ Dept.\ of Physics and Astronomy, Rutgers
University \par $\;\;\;$ P.O. Box 849, Piscataway, NJ 08855-0849
\vskip 1mm
\noindent [1] N.~Lehto and S.~\"Oberg, Phys. Rev. Lett. {\bf 80}, 5568 (1998).\par
\noindent [2]~J.~Bennetto, R.~W.~Nunes, and D.~Vanderbilt,
Phys. Rev. Lett. {\bf 79}, 245 (1997).\par
\noindent [3]~P.~N.~Keating, Phys. Rev. {\bf 145}, 637 (1966).\par
\noindent [4]~G.-X.~Qian and D.~J.~Chadi,
J. Vac. Sci. Technol. B {\bf 4}, 1079 (86).\par
\end{document}